# Effect of Trap States, Ion Migration, and Interfaces on Carrier Transport in Single-Crystal, Polycrystalline, and Thick Film Devices of Halide Perovskites $CH_3NH_3PbX_3$ (X = I, Br, Cl)

Mohd Warish, Gaurav Jamwal, Zara Aftab, Nidhi Bhatt, and Asad Niazi*

Read Online

ACCESS | Metrics & More | Article Recommendations | Supporting Information

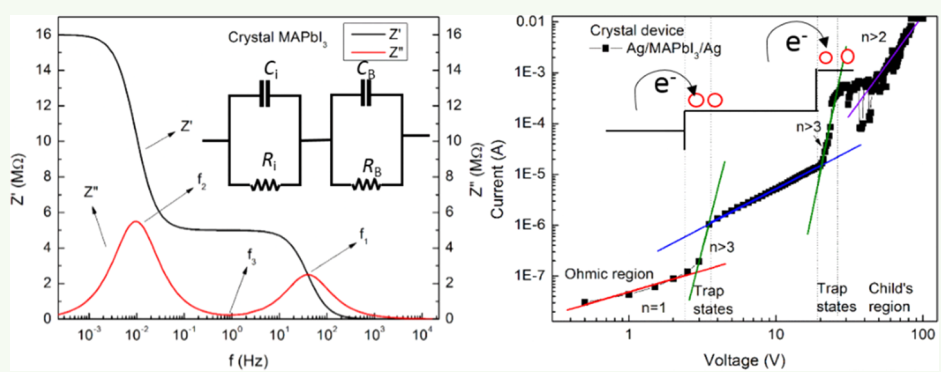

**ABSTRACT:** The understanding of the mixed ionic–electronic nature of charge transport in metal halide perovskites (MHPs) and the role of morphological and interface defects is crucial for improving the performance of MHP-based photovoltaic devices. We present the results of a parallel study on $MAPbX_3$ (X = I, Br, Cl), synthesized as solution-processed polycrystalline powders and as single crystals grown by a facile low-temperature-assisted technique. We have studied ionic–electronic charge transport in single-crystal and polycrystalline (pressed pellet and thick film) samples in order to compare the effect of defects and trap states associated with halide ion migration, device morphology, and interfaces at grain boundaries as well as at electrodes. The mobility of halide ions and associated Coulomb capture of electrons/holes was determined by dielectric and space charge limited current (SCLC) dark I–V measurements and also simulated using an ionic–electronic model. The defect capture cross section of electronic charge was found to be proportional to the simulated halide ion density $N_{ion}$, which varied in the range of $10^{16}$–$10^{22}$ cm$^{-3}$ depending on the halide ion. The trap state density from I–V measurements, $N_{trap} \sim 10^9$ to $10^{10}$ cm$^{-3}$, was found to be lower than those of previous reports. Single-crystal $MAPbI_3$ devices exhibited a low capture cross section ($\sigma_- \sim 10^{-16}$ cm$^{-2}$), high mobility ($\mu \sim 196$ cm$^2$/V-s), and large diffusion length ($L_D \sim 6$ μm). The study shows that nonradiative energy loss and carrier trapping are suppressed and transport properties are enhanced by reducing grain boundary effects, along with interface engineering to prevent halide ion accumulation at the electrodes.

## 1. INTRODUCTION

Metal halide perovskite (MHP) $MAPbX_3$ (MA= $CH_3NH_3^+$, X= I, Br, Cl) based materials have emerged as potential candidates for affordable photovoltaic energy generation and optoelectronic devices utilizing their charge carrier properties,[1−3] such as high absorption coefficients, long diffusion lengths over micrometer range, long charge carrier lifetime, high carrier mobility, and a high tolerance to defects.[4−7] The power conversion efficiency (PCE) for perovskite solar cells had reached 25.7% as of June 2022,[8] while tandem solar cells (perovskite/silicon) achieved 32.5% PCE, greater than the 26.1% PCE of pure silicon solar cells.[9,10] MHPs are being used in various devices, such as solar cells, photodiodes, lasers, photocatalysts, electrical diodes, and light-emitting diodes.[11−14]

Most MHP solar cells employ photoabsorbers in the form of polycrystalline thin films, in which the grain boundaries, voids, and surface defects act as recombination sites, leading to significant efficiency losses. The grain boundaries act as defect centers to create electron trap states arising from Coulomb capture, which decrease the open circuit voltage and reduce the photovoltaic performance. Polycrystalline absorber materials

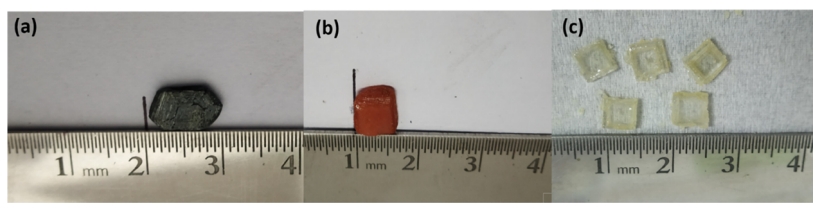

**Figure 1.** Halide perovskite crystals grown by the low-temperature-assisted technique: (a) MAPbI$_3$, (b) MAPbBr$_3$, and (c) MAPbCl$_3$.

also suffer from poor long-term stability issues. There has been an increasing effort in studying absorber materials in the form of large single crystals. Single crystals of MHP are more stable toward degradation and exhibit better optoelectronic properties as compared to polycrystalline materials due to fewer grain boundaries and other defects.[4,15] They exhibit high absorption coefficients ($10^5$ cm$^{-1}$), long diffusion lengths (>10 $\mu$m), long charge carrier lifetimes (up to microsec), high carrier mobility (up to $10^3$ cm$^2$/V-sec), and a high tolerance to defects with very low trap state density ($10^9$ cm$^{-3}$), resulting in higher power conversion efficiency.[15,16] Various methods of growing millimeter-sized MHP crystals have been reported, such as seed growth, ion-convection-based top-seed solution growth, antisolvent vapor diffusion process, ultralow cooling rate at low temperatures, and rapid crystallization by inverse temperature crystallization.[17−19]

In the present work, we report a comparative study on MAPbX$_3$ (X = I, Br, Cl) solution-processed polycrystalline powders, as well as crystals synthesized by a low-temperature-assisted solution growth method at concentrations lower than the saturation concentration typically required for crystallization.[15]

The halide ions in MHP materials exhibit a significant degree of mobility in the crystal lattice. These mobile ions and the associated vacancies, along with polycrystalline grain boundaries, lead to a number of detrimental effects such as nonradiative recombination, carrier trapping, formation of potential barriers at interfaces, and imbalance of ionic−electronic charge carrier concentration. Understanding and suppressing the above ionic effects are crucial for enhancing device performance. However, the carrier dynamics in these semiconductors is still not well understood due to the mixed ionic−electronic character as well as variations in morphology and grain size, which affect the charge carrier properties and device performance.[20,21]

We have studied the effect of different morphologies and the mixed nature of ionic−electronic charge carrier properties on thick film, polycrystalline bulk, and single-crystal samples characterized via structural, optical, dielectric, and space charge limited current (SCLC) dark I−V measurements. The UV−vis and photoluminescence (PL) spectroscopy indicate a very small defect concentration for all samples, as evidenced in the low absorptive energy loss of a few meV. The dielectric measurements establish the role of grain boundaries in charge carrier trapping, with a low defect capture cross section observed for single crystals. The polarizability and density of the mobile halide ions were also observed to influence the dielectric response, with the Cl samples exhibiting a greater probability of charge capture. The calculated impedance was compared with simulations based on an ionic−electronic model to extract ionic parameters such as density, Debye length, diffusion coefficient, and mobility. The SCLC dark I−V measurements show improved transport properties such as greater charge carrier mobility and low trap state densities compared to previous studies. The effective density of mobile halide ions ($N_{eff}$), obtained from the difference between the simulated ionic density ($N_{ion}$) and the trap density ($N_{trap}$) from SCLC measurements, was found to vary over a large range ($10^2$−$10^{12}$ cm$^{-3}$) depending on the halide ion and device morphology, significantly influencing their electrical transport and optoelectronic properties.

To the best of our knowledge, this is the first combined study of I, Br, and Cl MHP polycrystalline and single-crystal samples to understand the mixed ionic−electronic movement across the grain boundaries and absorber−electrode interfaces.

## 2. EXPERIMENTAL SECTION

**2.1. Synthesis.** The following chemicals were used as supplied for the synthesis of the MAPbX$_3$ samples: CH$_3$NH$_2$ (40% aq. soln.), PbI$_2$ (Alfa Aesar, 99%), PbBr$_2$ (Loba Chemie, 99%), PbCl$_2$ (Alfa Aesar, 99%), HI (CDH, 54%), HBr (CDH, 40% extra pure), HCl (Loba Chemie, 35%), γ-butyrolactone (GBL) (Loba Chemie, 99%), N,N-dimethylformamide (DMF) (CDH), and dimethyl sulfoxide (DMSO) (Loba Chem, 99%). The precursor CH$_3$NH$_3$I was synthesized by dropwise adding 30 mL of HI to 21 mL of CH$_3$NH$_2$ in a stirring flask kept in an ice bath under a constant flow of N$_2$. The transparent solution formed after 2 h of stirring was dried at 60 °C for 24 h. The precipitate thus obtained was washed thoroughly in diethyl ether and dried at 70 °C for 12 h, resulting in a snowy white powder of CH$_3$NH$_3$I. CH$_3$NH$_3$Br and CH$_3$NH$_3$Cl were synthesized using the same procedure as above, using HBr and HCl, respectively.

Polycrystalline CH$_3$NH$_3$PbI$_3$ was prepared by dissolving stoichiometric amounts of as-prepared CH$_3$NH$_3$I and PbI$_2$ in 10 mL of GBL solvent, which was then stirred for 6 h at ∼95 °C until a transparent solution was obtained. The solution was filtered and dried in an oven at ∼100 °C. The blackish polycrystalline powder of CH$_3$NH$_3$PbI$_3$ was obtained after about 24 h. A similar synthesis procedure was used for CH$_3$NH$_3$PbBr$_3$ and CH$_3$NH$_3$PbCl$_3$, using DMF and DMSO as respective solvents (Supporting Information (SI), Section S2).

Single crystals of methylammonium lead halide (CH$_3$NH$_3$PbX$_3$, X = I, Br, Cl) were grown using a low-temperature-assisted solution growth method, described below.

The precursors CH$_3$NH$_3$I and PbI$_2$ were separately dissolved in 5 mL of GBL solvent each, with stirring for 2 h at 80 °C. This was required as combining the precursors before dissolution resulted in unreacted precipitates, observed during the polycrystalline powder synthesis. The precursor solutions were then combined and stirred for 4−6 h at 80 °C. The homogenized solution was filtered to obtain a transparent solution of CH$_3$NH$_3$PbI$_3$. This solution was then used as the mother liquor for growing single crystals as well as to prepare a thick film device.

Prior to crystal growth, the solution was initially stabilized by keeping it in an oven at 60 °C for 24 h. The temperature was then increased to 90−95 °C, and black crystals of 5−10 mm size were obtained after 24 h (Figure 1). The crystals of CH$_3$NH$_3$PbBr$_3$ and CH$_3$NH$_3$PbCl$_3$ were synthesized using a similar protocol at different synthesis temperatures and using appropriate solvents (SI, Section S2).

In order to obtain high-quality crystals, it is critical to have a proper concentration of the seed solution as well as an optimal growth temperature. A low concentration of the solution (<1 molar) is insufficient to initiate seeding, while using 2 molar or greater concentration has been reported to cause defects due to rapid crystal growth [15]. In the present study, 1 molar concentration was used to grow high-quality crystals of MHPs at temperatures below the normal threshold for rapid crystal growth. The best growth of CH$_3$NH$_3$PbI$_3$ crystals was obtained at a temperature of 90 °C, while exceeding 100 °C induced defects, as well as breakdown of the compound into its constituents.

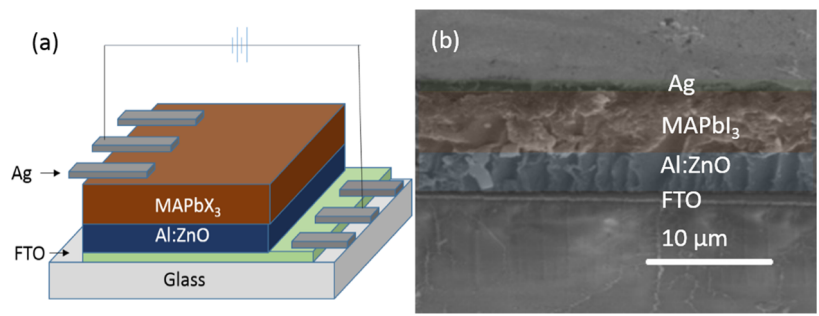

**Figure 2.** (a) Schematic diagram for device architecture and (b) cross-sectional SEM image of electron-assisted thick film device of MAPbI$_3$.

**2.2. Device Fabrication.** Fluorinated tin oxide (FTO) conducting glass (1.5 cm × 1.5 cm × 1.1 mm) with a 135 ± 15 nm FTO coating thickness and a surface resistance of 15 Ω/square was used as a substrate. The substrate was etched along three edges using Zn powder and 2 molar HCl, leaving one edge for deposition of the bottom electrode. The etched FTO was washed gently with soap water, sonicated in deionized water followed by acetone, and then dried using nitrogen. For making the electron-type device, the substrate was masked using thermal masking tape and preheated at 200 °C for 10 min before spin-coating with Al: ZnO (AZO) nanoparticles (Sigma-Aldrich, 2.5% viscosity 3.5 cP, work function 4.3 eV) at 800 rpm for 1 min, followed by 2500 rpm for 2 min. The coated substrate was annealed at 200 °C for 30 min. Once the substrate had cooled to below 100 °C, the absorber thick layer was deposited by spin-coating at 800 rpm for several minutes in the presence of nitrogen, followed by annealing at 70 °C for 10 min. The Ag electrode was deposited using thermal evaporation. The active area of the device was 0.7 cm × 0.7 cm. The device is shown schematically in Figure 2a. The thickness of the absorber layer was determined by cross-sectional SEM and HR-SEM imaging to be ∼5 μm (Figure 2b). For the hole-type device, the substrate was kept at room temperature while spin-coating PEDOT:PSS (Sigma-Aldrich, 2.8 wt % in aq dispersion), used as a hole transporting layer. After the deposition of PEDOT:PSS, the substrate was annealed at 70 °C, followed by deposition of the absorber layer and Ag contacts as before.

## 3. CHARACTERIZATION TECHNIQUES

The structural parameters of MAPbX$_3$ (X= I, Br, Cl) were extracted from powder XRD measurements on crushed single-crystal samples in the range of 2θ = 10−60° on a Rigaku Ultima IV Diffractometer with a Cu Kα X-ray source. Thermogravimetric analysis of polycrystalline samples using a Setaram Instrumentation, LABSYS EVO 1150 °C DSC131 EVO analyzer, was used to confirm the stability of the samples until 100 °C. The absorption spectra of polycrystalline samples were obtained in reflectance mode using an Agilent Cary Eclipse Spectrometer equipped with a microsecond xenon flash lamp. The photoluminescence (PL) and time-resolved photoluminescence (TRPL) spectra were obtained on solutions. The PL spectra were recorded by using a Comspec M550 UV−vis spectrometer. The TRPL spectra were recorded on a Horiba-Delta Flex 01-DD spectrometer using a time-correlated single photon counting (TCSPC) technique using a pulsed diode laser source of 529 nm for I and 401 nm for Br and Cl, respectively, with the fluence kept at 0.1−0.3 μJ/cm$^2$. The morphological characterization of single crystals, polycrystalline pellets, and thick film device cross sections was carried out using a field emission scanning electron microscope (FE-SEM), JEOL/7610F-FE-SEM. A high-resolution transmission electron microscope (HR-TEM), JEOL/JEM-F200, with Cu grid (CF200-CU, Carbon film 200 mesh) was used to image the nanoparticles and characterize the distinct *hkl* planes for MAPbX$_3$ (X= I, Br, Cl). The dielectric data of single-crystal and polycrystalline sintered pellets were obtained by using an Agilent E4980A precision LCR meter. The dark current−voltage (I−V) measurements were carried out using a Keithley 2400 sourcemeter, using the space charge limited current (SCLC) method.

## 4. RESULTS AND DISCUSSION

**4.1. X-ray Diffraction.** Figure 3 shows the powder XRD pattern of crushed crystals of MAPbX$_3$ (X = I, Br, Cl) confirming

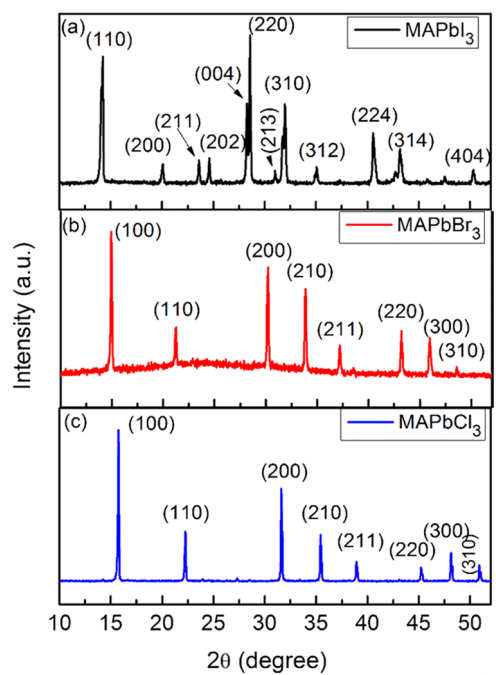

**Figure 3.** Powder XRD of crushed crystals of MAPbX$_3$ (X= I, Br, Cl) crystallizing in (a) *I4/mcm* tetragonal phases and (b, c) *Pm$\bar{3}$m* cubic phase.

well-formed *Pm$\bar{3}$m* cubic (X = Cl, Br) and *I4/mcm* tetragonal (X = I) crystalline phases. No residual peaks of PbX$_2$ (X = I, Br, Cl) were observed. The corresponding powder XRD pattern for polycrystalline samples is given in Figure S1. Table 1 lists the cell parameters obtained by the Rietveld refinement of the XRD data using Fullprof. The Rietveld refinement plots are shown in

Table 1. Structural Parameters of MAPbX$_3$ (X= I, Br, Cl) Obtained from XRD and SEM/HR-TEM

| MAPbX3 | phase | cell parameter (Å) | | grain size | nanoparticle size | |
|---|---|---|---|---|---|---|
| | | single crystal | polycrystalline | SEM (μm) | SEM (nm) | HR−TEM (nm) |
| I | tetragonal $I4/mcm$ | a = 8.8760(2)  c = 12.6810(5) | a = 8.8713(3)  c = 12.6546(8) | 10 | 10 | 15 |
| Br | cubic $Pm\bar{3}m$ | 5.9204(1) | 5.9285(1) | 10 | 10 | 10 |
| Cl | cubic $Pm\bar{3}m$ | 5.6904(3) | 5.6841(6) | 5 | 5 | 6 |

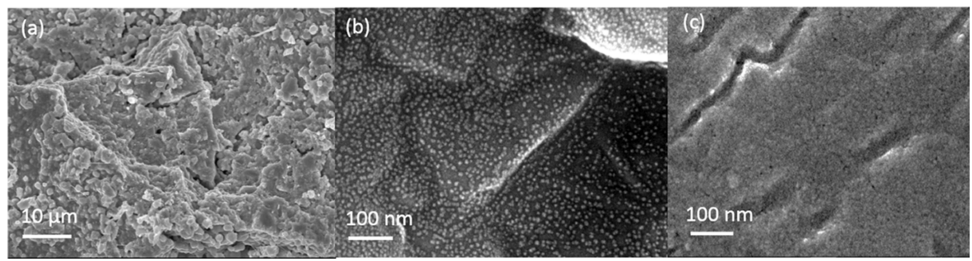

**Figure 4.** SEM images of MAPbI$_3$ (a, b) polycrystalline pellets and (c) single crystals.

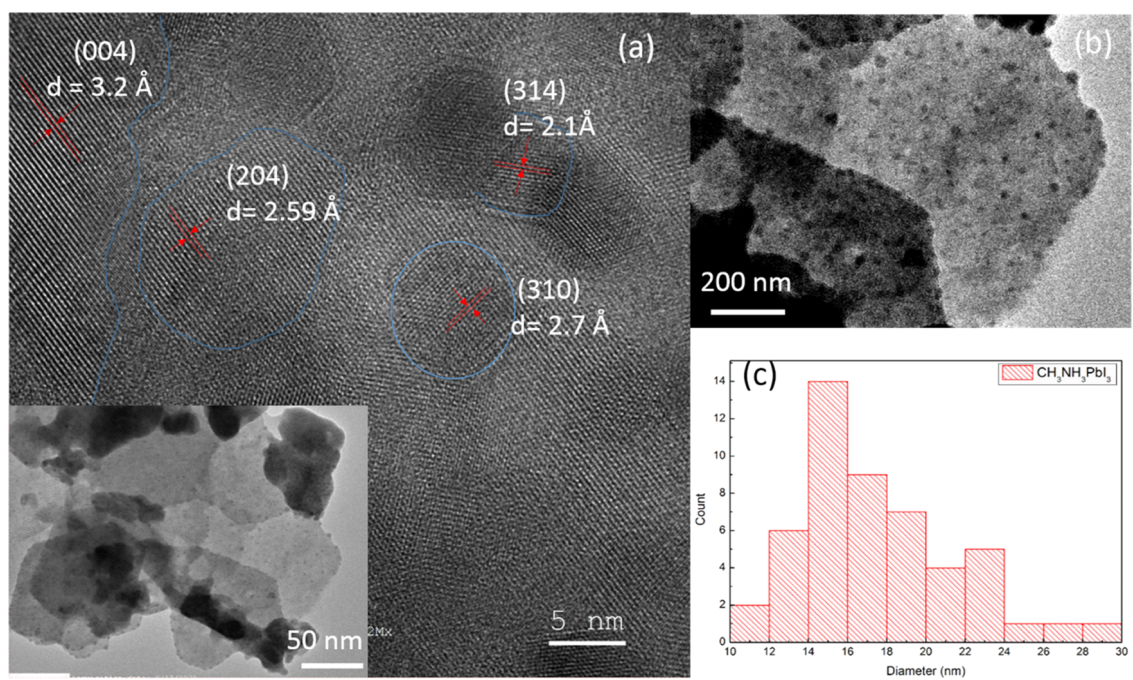

**Figure 5.** HR-TEM images of MAPbI$_3$ showing (a) $d(hkl)$ planes, (b) nanoparticles, and (c) a histogram of nanoparticle size distribution.

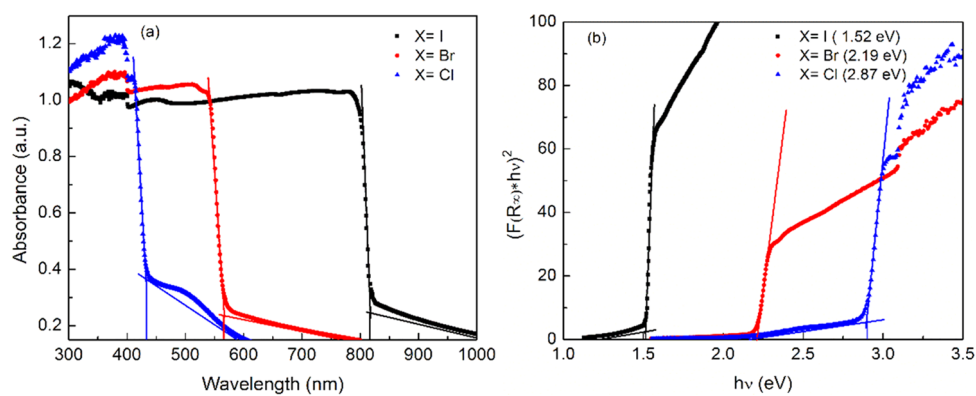

**Figure 6.** Absorption spectra for MAPbX$_3$ (X= I, Br, Cl): (a) absorption spectra and (b) Tauc plot.

Figure S2. The lattice parameters compare well with literature.[15,22,23]

### 4.2. SEM and HR-TEM. 

Figure 4 shows the SEM micrographs of MAPbI$_3$ (a, b) polycrystalline pellets and (c) single crystals. The samples exhibit a dual morphological character: large grains of size ∼1 to 10 μm, with nanoparticles of ∼5 to 10 nm on their surface. Figure 5 shows the TEM images of MAPbI$_3$ at 5 and 20 nm resolutions. The samples were prepared by the ultrasonic dispersion of crushed crystals in isopropyl alcohol (IPA) for ∼20 min, followed by drop casting the dispersed suspension on a TEM copper grid. The TEM images were analyzed using ImageJ software to extract the d($hkl$) plane spacing and size distribution histograms. The results are given in Table 2. Nanoparticles of an average size of 5 to 15 nm can clearly be observed on the grain surfaces. The size of the nanoparticles matches that of the SEM images (Figure 4). The d($hkl$) plane spacing matches well with that obtained from powder XRD data, indicating that the nanoparticles are locations of nucleation and growth during the crystallization process. The corresponding SEM and HR-TEM scans of the Br and Cl samples are given in the SI (Figures S3 and S4).

### 4.3. UV−Visible Spectroscopy. 

Figure 6a shows the absorbance of MAPbX$_3$ (X = I, Br, Cl) crushed crystals obtained from the spectra measured in reflectance mode. The absorption onsets show a clear band edge cutoff with no excitonic signals or absorption tails, indicating a low defect concentration.[15,24,25] The absorption onset values of 834 nm for I, 567 nm for Br, and 441 nm for Cl correspond well with the colors of the as-grown crystals: black (I), orange (Br), and transparent (Cl), which are shown in Figure 1.

The optical band gap was determined from the Tauc plot, $(F(R\infty)*h\nu)^{1/m}$ vs ($h\nu$), shown in Figure 6b, where $m = 1/2$ for direct band-gap semiconductors. The band-gap values obtained from the Tauc plot are slightly underestimated due to excitonic absorption at the band edge within the sample, as indicated in Figure 6b. We estimate that the actual absorption onset and correspondingly the band gap would be about 0.1 eV greater than that obtained from the Tauc plot.[26] The results are summarized in Table S1. The estimated values of the optical band gap, 2.87 eV (Cl), 2.19 eV (Br), and 1.52 eV (I), are comparable with other published works on bulk powders but lower than those reported for thin films.[15,19,27,28]

### 4.4. Photoluminescence. 

Figure 7 shows the photoluminescence (PL) spectra of MAPbX$_3$ (X = I, Br, Cl) with peaks at about 732, 474 (±10), and 440 nm for I, Br, and Cl, respectively. The PL peaks of Br and I show a blue shift in comparison to the absorption onset values, indicating lower trap density as compared to that reported by previous studies.[27] For MAPbCl$_3$, the values of the absorption onset (434 nm) and Pl peak (440 nm) are nearly coincident, indicating Rayleigh emission with no excitonic signature or absorption tails, which can be attributed to negligible defect states.[15,27]

### 4.5. Time-Resolved Photoluminescence. 

Figure 8 shows the time-resolved photoluminescence (TRPL) spectra of the

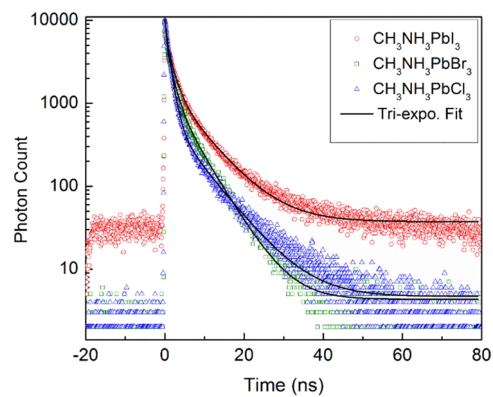

**Figure 8.** Time-resolved photoluminescence (TRPL) for MAPbX$_3$ (X = I, Br, Cl). The solid lines are fits using the triexponential function of eq 2.

perovskite absorbers in respective solvents. The luminescence spectra contain contributions from charge carrier recombination occurring through multiple pathways with varying lifetimes— direct band-to-band radiative, trap-assisted radiative, and excitonic recombination processes.[28,29] The trap-assisted recombination is associated with nonradiative ultrafast (sub-picosecond) thermalization of the excited electrons into trap states with short lifetimes.[30−32]

The TRPL decay was fitted by a triexponential decay function (eq 1)

$$y = Ae^{-t/\tau_1} + Be^{-t/\tau_2} + Ce^{-t/\tau_3} \quad (1)$$

where $\tau_1$, $\tau_2$, and $\tau_3$ are the average lifetime decay constants associated with trap-assisted radiative, direct band-to-band radiative, and excitonic recombination processes, respectively.[33] The average decay time $\tau_{avg}$ was calculated by the weighted contribution of the recombination routes described above (eq 2)

$$\tau_{avg} = \frac{A\tau_1^2 + B\tau_2^2 + C\tau_3^2}{A\tau_1 + B\tau_2 + C\tau_3} \quad (2)$$

The detailed time-resolved PL parameters for I, Br, and Cl are listed in Table S1. The $\tau_{avg}$ was about 1 ns for all compositions.

The fast decay time ($\tau_1$) for MAPbX$_3$ (X = I, Br, Cl) was determined to be 2.26 ns (I), 1.39 ns (Br), and 1.2 ns (Cl) and can be associated with trap-assisted recombination. In comparison, the longer decay time ($\tau_2$), 10 ns (I), 5 ns (Br), and 7 ns (Cl), can be ascribed to direct band-to-band recombination. The radiative recombination rate was found to be directly proportional to the ionic and carrier concentration obtained from the simulation study based on an ionic model as well as from SCLC measurements (Tables S5 and S6).[34,35]

### 4.6. Dielectric Study. 

Dielectric measurements on MAPbX$_3$ (X = I, Br, Cl) single crystals (1 mm) and polycrystalline pellets (1 mm thick, sintered at 90 °C) were performed in the frequency range of 20 Hz–2 MHz at 300 K, using Ag paste contacts. The thickness of the MAPbX$_3$ dielectric used in the capacitor allows the study of ionic and charge carrier dynamics in the bulk material as compared with earlier reports on films less than 1 $\mu$m thick. The frequency range corresponds to the space charge-induced dielectric response for both ionic and electronic conductions. Temperature-dependent spectra from 100 to 400 K were recorded to observe the effect of phase symmetry and dynamical disorder of varied grain sizes and grain boundaries. The results of polycrystalline samples are shown in Figures 9 and S5, while those of single crystals are given in Figures S6 and S7.

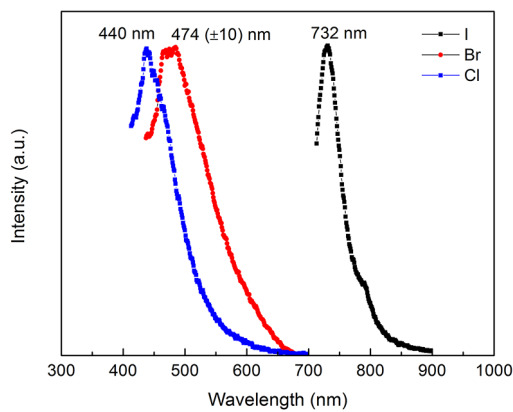

**Figure 7.** Photoluminescence spectra for MAPbX$_3$ (X = I, Br, Cl).

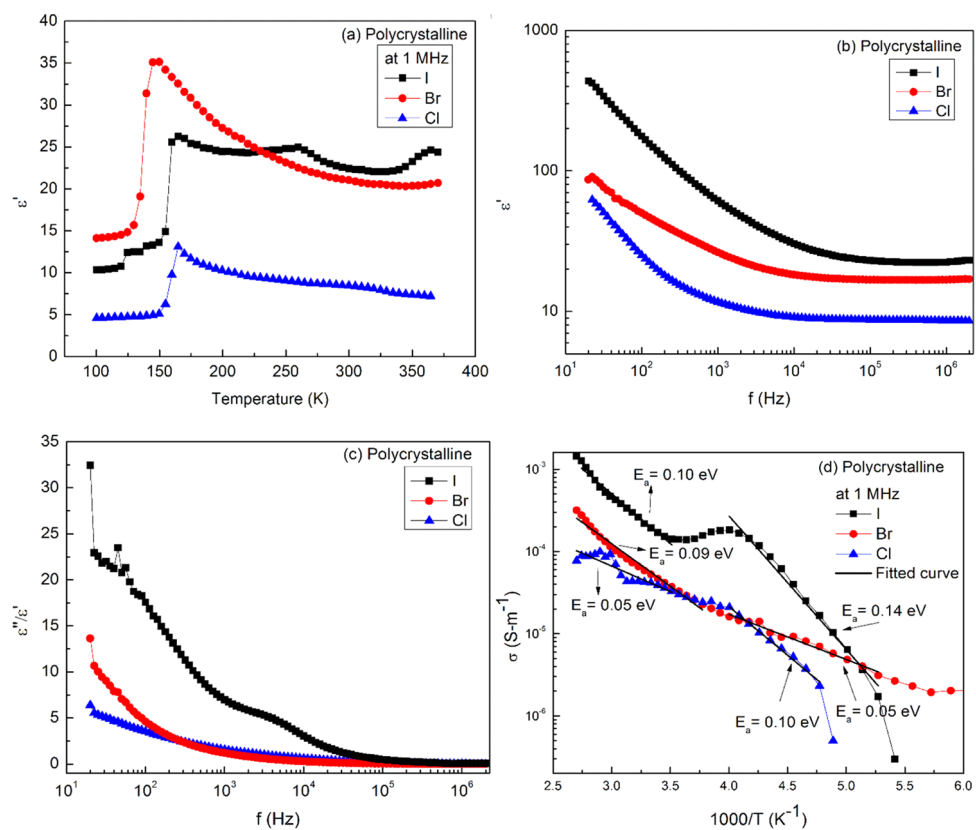

**Figure 9.** (a) Dielectric constant vs temperature of polycrystalline MAPbX$_3$ (X = I, Br, Cl), (b) dielectric constant vs frequency, (c) dissipation factor vs frequency, and (d) Arrhenius plot of conductivity vs temperature at 1 MHz.

The dielectric constant ($\varepsilon'$) was calculated by (eq 3)

$$\varepsilon' = Cd/\varepsilon_o A \qquad (3)$$

where $C$ is the capacitance of the material, $d$ is the separation between the parallel metallic plates of Ag, $\varepsilon_o$ is the permittivity of free space, and A is the area of parallel plates.

Figures 9 and S7 show the frequency-dependent dielectric response. The results are summarized in Table S2. The $\varepsilon'$ decreases with an increase in frequency for all samples. The low-frequency values of $\varepsilon'$ are higher in the case of single crystalline capacitor samples as compared to those of the corresponding polycrystalline ones. This could be due to fewer defects and larger grain size in the former. The ionic conductivity (Figures 9d, S5b, and S6b,d) was calculated using the relation $\sigma_i = \varepsilon_o \varepsilon' \omega \tan \delta$, where $\varepsilon_o$ and $\varepsilon'$ are the permittivity of free space and dielectric constant, respectively, $\omega$ is the angular frequency, and tan $\delta$ is the loss factor.[36]

Figures 9, S5, and S6 show the temperature variation of the dielectric response at 100 Hz and 1 MHz in the range of 100–400 K, measured while heating. The orthorhombic–tetragonal structural transitions in the temperature range of 130–170 K can be observed in the sharp discontinuities in the dielectric constant in both polycrystalline and single crystals and match with previously reported studies.[37] The tetragonal–cubic structral transitions at higher temperatures are not as sharply defined. The sharp increase in low-frequency $\varepsilon'$ at high temperatures could be a signature of movement/relaxation of halide ions and space charge carriers as reported previously.[21,38]

In Figures 9a–d and S5, the polycrystalline MAPbI$_3$ sample shows anomalous temperature-dependent behavior of the dielectric constant $\varepsilon'(T)$ and associated conductivity $\sigma(T)$ in the temperature range of 250–300 K with a negative slope. The behavior is more pronounced in polycrystalline samples (Figures 9 and S4) as compared to the single crystals (Figure S6) in which a negative slope is seen from about 300–325 K. MAPbI$_3$ is known to undergo a structural phase transformation at about 330 K from (high-temperature) cubic $Pm\bar{3}m$ to (low-temperature) tetragonal $I4/mcm$. We suggest that the anomaly in $\varepsilon'(T)$ is due to fluctuations in the dipolar response of the MA and the PbI$_6$ octahedra in the crystal lattice that build up in the vicinity of the tetragonal–cubic structural phase transformation. The large response of the polycrystalline pressed pellets as compared to that of single-crystal samples could be due to the disorder due to the small grain size ($\sim$10 $\mu$m) of the former. The anomalous dielectric behavior is not observed in the Br and Cl samples. We suggest that the larger size of the I$^-$ ion as compared to that of Br$^-$ and Cl$^-$ leads to a larger dipolar response. A conclusive explanation of the observed dielectric response anomaly would require a detailed study of the bond-vibrational dynamics of polycrystalline and single-crystal MHP samples, which is beyond the scope of the current work.

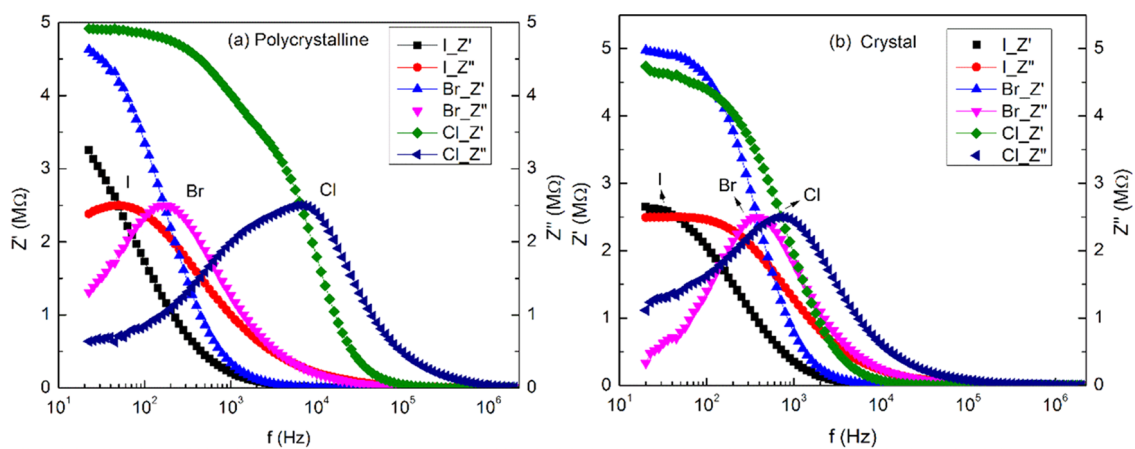

**Figure 10.** Frequency dependence of impedance in MAPbX$_3$ (X = I, Br, Cl); (a) polycrystalline and (b) single crystals.

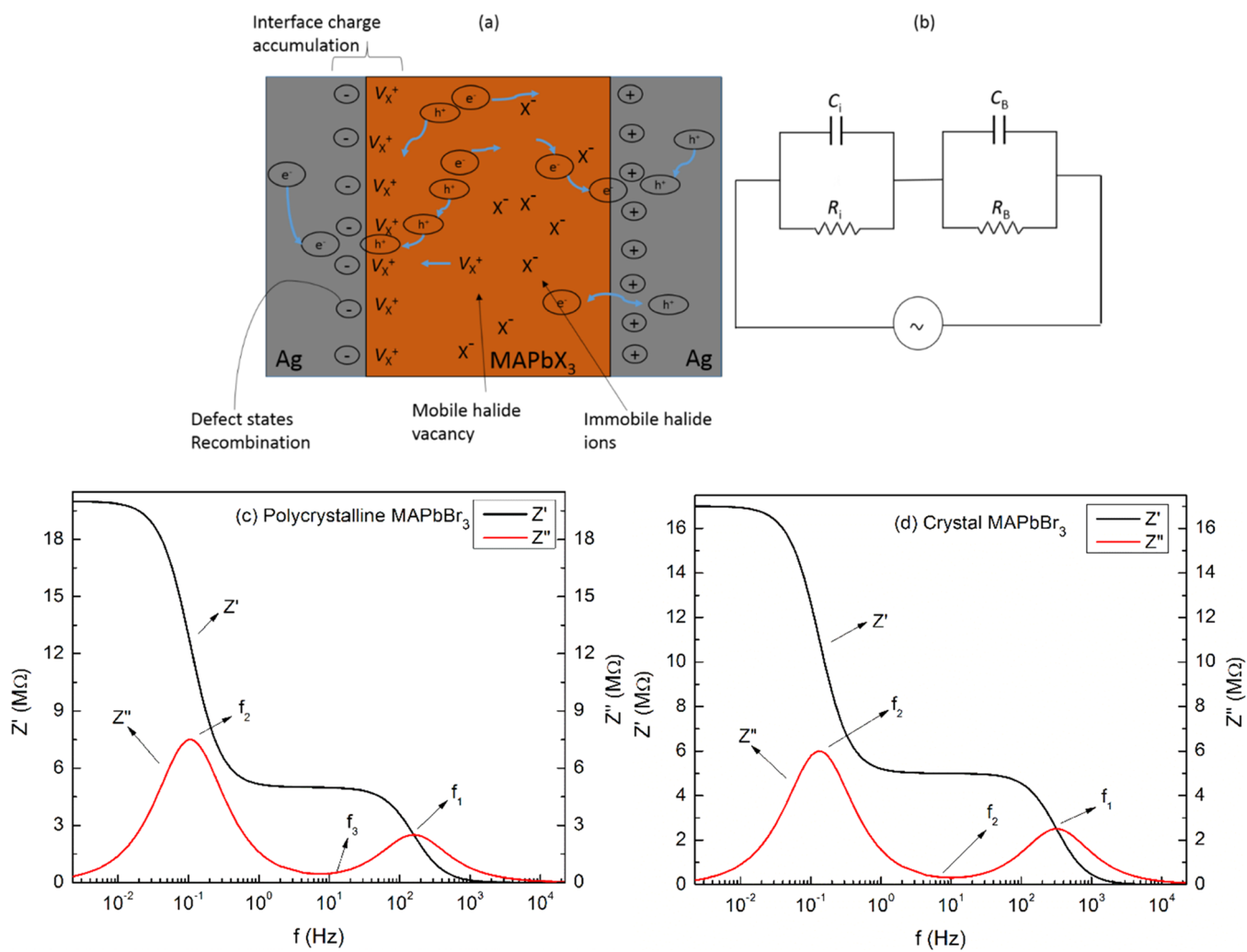

**Figure 11.** (a, b) Schematic representation of ion and charge carrier movement and accumulation at grain boundaries and interface, and (c, d) simulated impedance curves for MAPbBr$_3$ using the parallel RC circuit model.

Table S2 lists the activation energy ($E_a$) at 100 Hz calculated from the Arrhenius plot (Figures 9d, S5b, S6b,d), $\sigma = \sigma_0 \exp(-E_a/k_B T)$, where $\sigma$ is the ionic conductivity, $\sigma_0$ is the static conductivity, $k_B$ is the Boltzmann constant, and $T$ is the temperature. For polycrystalline samples, $E_a$ was found to be 0.15 eV (I), 0.13 eV (Br), and 0.10 eV (Cl), which is comparable to that of ionic conductors, 0.55 eV for Cs$_3$H(SeO$_4$)$_2$ and 0.49 eV for Pb$_3$H(SeO$_4$)$_2$, respectively.[39] However, at 1 MHz, the $E_a$ was much lower at 98 meV (I), 90 meV (Br), and 50 meV (Cl). The activation energy at low frequency for the single-crystal samples shows a similar trend.

It is important to explore the role of defects such as ion vacancies and grain boundary interfaces in order to reduce nonradiative losses and to optimize charge carrier transport. The high polarizability of MAPbX$_3$ and the mobility of the halide ion strongly influence the space charge-induced local electric fields within crystalline grains. This affects charge carrier transport across grain boundaries, which act as trapping sites. The electronic charge is captured due to the Coulomb attraction by positive defects in its path, with the defect capture cross section ($\sigma_-$) described by eq 4[21]

$$\sigma_- = \frac{q^4}{16\pi\varepsilon_o^2\varepsilon_r^2 k_B^2 T^2} \quad (4)$$

where $q$ is the elementary charge, $\varepsilon_o$ is the vacuum permittivity, $\varepsilon_r$ is the dielectric constant, $k_B$ is Boltzmann's constant, and $T$ is the temperature, and where the Coulomb potential energy equals the thermal energy ($k_B T$).

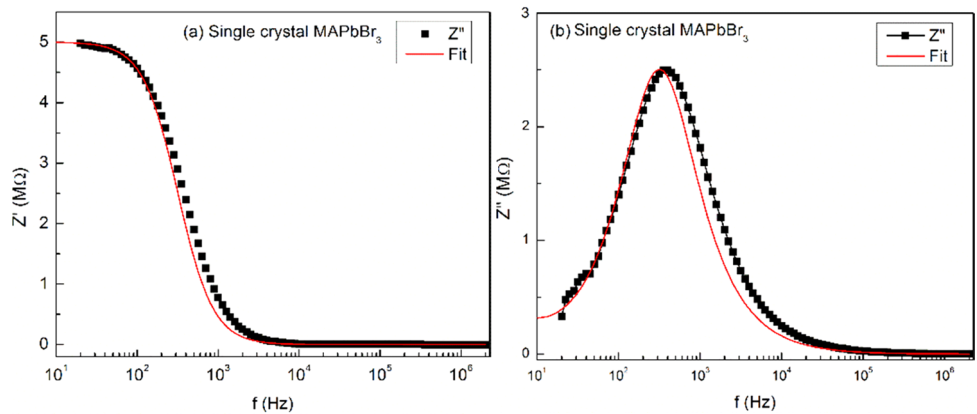

**Figure 12.** (a, b) Impedance of MAPbBr$_3$ crystals fitted by the equivalent circuit model, showing the bulk relaxation region ($f_1$).

Table S3 lists the dielectric constant $\varepsilon'$ in the static ($\varepsilon_s$, 20 Hz) and dynamic ($\varepsilon_{inf}$, 1 MHz) limits and the corresponding defect capture cross section ($\sigma_-$) for polycrystalline and single-crystal MAPbX$_3$ (X = I, Br, Cl). The Br and Cl single crystals show lower $\sigma_-$ values as compared to the polycrystalline samples, indicative of a lower density of defect states in the former. On the other hand, the I sample showed a similar response in the polycrystalline as well single-crystal samples. The static $\sigma_-$ of I was also much smaller than those of the Br and Cl samples. These differences are possibly due to the difference in ionic size and electronegativity of the halide ions.

Figure 10 shows the real ($Z'$) and imaginary ($Z''$) impedances in the frequency range of 20 Hz–2 MHz determined from the dielectric measurements on polycrystalline and single-crystal MAPbX$_3$ parallel plate capacitors. The $Z'$ for polycrystalline MAPbI$_3$ and MAPbBr$_3$ becomes negligible at frequencies above $10^3$ Hz, while for MAPbCl$_3$, it remains large until $10^5$ Hz. The relaxation peak (maxima) $f_1$ for $Z''$ was observed at 48, 175, and 7000 Hz and the expected minima ($f_3$) at 0.41, 6.46, and 12 Hz for I, Br, and Cl, respectively. For single-crystal capacitors, the relaxation peak (maxima) $f_1$ was observed at 90 Hz (I), 383 Hz (Br), and 743 Hz (Cl) and the minima ($f_3$) at 1 Hz (I), 8.81 Hz (Br), and 2.08 Hz (Cl).

The mixed electronic–ionic charge carrier dynamics has been analyzed using a model for ionic mobility and density in liquids and electrolytes.[40] In this model, the negatively charged halide ions are considered fixed, and the positively charged halide ion vacancies are mobile (Figure 11a). Under an external ac electrical bias of frequency $f$, the ion vacancies move to the cathode and form a layer at the interface. A Helmholtz double layer is developed that behaves as a parallel $R_i C_i$ circuit, where $C_i$ and $R_i$ are the capacitance and resistance of the ionic layer. This $R_i C_i$ element acts in series with the parallel $R_B C_B$ circuit corresponding to the bulk material,[39] where $C_i \gg C_B$, since the thickness of the Helmholtz layer is much smaller than the bulk dielectric (Figure 11a,b). The total series impedance of the two RC circuits can be calculated by eqs 5 and 6

$$Z' = \frac{R_B}{1 + R_B^2 \omega^2 C_B^2} + \frac{R_i}{1 + R_i^2 \omega^2 C_i^2} \quad (5)$$

$$Z'' = -\left(\frac{R_B^2 \omega C_B}{1 + R_B^2 \omega^2 C_B^2} + \frac{R_i^2 \omega C_i}{1 + R_i^2 \omega^2 C_i^2}\right) \quad (6)$$

The estimated time constants for the parallel RC circuit, extracted from the experimental data using the above ionic model, are expressed by eq 7

$$t_1 = \frac{1}{2\pi f_1(=\omega_1)} = R_B C_B, \quad t_2 = \frac{1}{2\pi f_2(=\omega_2)} = R_i C_i \quad (7)$$

The relaxation time constant $t_1$ for the bulk space charge can be calculated by using $Z' = R_B/2$ for the frequency range $\omega > 1/R_i C_i$. The relaxation time $t_2$ of the space charge in the ion vacancy Helmholtz layer corresponds to the low-frequency region $\omega < 1/R_B C_B$, where $Z' = R_B + (R_i/2)$. Figures 11c,d and S8 show the simulated impedance curves with peaks at frequencies corresponding to the relaxation of the bulk ($f_1$) and the Helmholtz layer ($f_2$). The minimum between the two relaxation peaks for $Z''$ at frequency ($f_3$) corresponds to the relaxation of the space charge at the interface between the bulk and the Helmholtz layer with a time constant ($t_3$) given by eqs 8 and 9

$$t_3 = \frac{1}{2\pi f_3(=\omega_3)} = \frac{1}{2\pi f_1(=\omega_1)}\sqrt{\frac{C_i}{C_B}} = t_1\sqrt{\frac{L}{\lambda_d}} \quad (8)$$

$$D_{ion} = \frac{\lambda_d^2}{t_3} \quad (9)$$

where $L$ is the thickness of the sample, $\lambda_d$ is the Debye length for ionic diffusion at the interface ($L_{ion} = \sqrt{D_{ion}\tau}$), and $D_{ion}$ is the simulated ion diffusion constant in the lower-frequency range.[39]

Figure 12 shows the experimental impedance data for the single-crystal Br sample fitted by the model described in eqs 5 and 6. The model provides a close fit to the bulk relaxation maximum ($f_1$) at the high frequencies observed experimentally.

Table 2. Simulated Ionic Transport Parameters

| Absorber | | Debye length $\lambda_d$ ($\mu$m) | Ionic diffusion coefficient $D_{ion}$ (cm$^2$/s) | Ionic mobility $\mu_i$ (cm$^2$/V–s) | Ionic density $N_{ion}$ (cm$^{-3}$) 100 Hz | 1 MHz |
|---|---|---|---|---|---|---|
| polycrystalline | I | 0.70 | $4.13 \times 10^{-8}$ | $1.65 \times 10^{-6}$ | $1.43 \times 10^{17}$ | $4.28 \times 10^{17}$ |
| | Br | 1.42 | $7.62 \times 110^{-7}$ | $3.00 \times 10^{-5}$ | $2.81 \times 10^{15}$ | $3.43 \times 10^{16}$ |
| | Cl | $3.03 \times 10^{-3}$ | $6.55 \times 10^{-12}$ | $2.62 \times 10^{-10}$ | $6.15 \times 10^{20}$ | $1.90 \times 10^{21}$ |
| crystal | I | 0.53 | $1.61 \times 10^{-8}$ | $6.46 \times 10^{-7}$ | $1.31 \times 10^{18}$ | $9.77 \times 10^{18}$ |
| | Br | 1.00 | $5.50 \times 10^{-7}$ | $2.20 \times 10^{-5}$ | $2.52 \times 10^{17}$ | $1.47 \times 10^{17}$ |
| | Cl | $1.63 \times 10^{-2}$ | $1.80 \times 10^{-11}$ | $7.20 \times 10^{-10}$ | $2.07 \times 10^{22}$ | $1.91 \times 10^{22}$ |

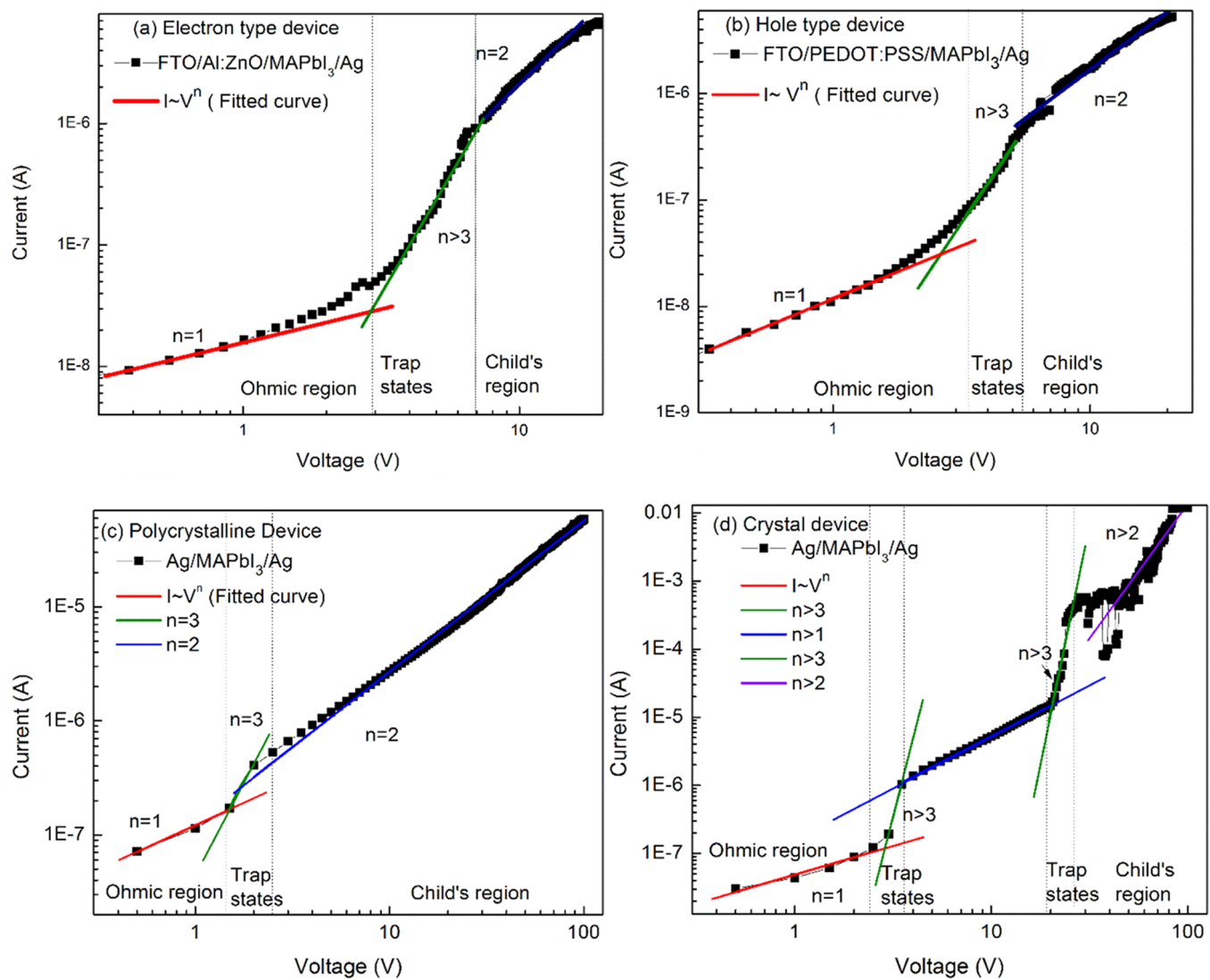

**Figure 13.** I−V curves for MAPbI$_3$ devices: (a) thick film electron-assisted, (b) thick film hole-assisted, (c) polycrystalline, and (d) single crystals.

The ionic parameters obtained from simulation are listed in Table 2. The Helmholtz layer relaxation time $t_2$ corresponding to the maximum ($f_2$) and the space charge relaxation time $t_3$ corresponding to the minimum ($f_3$) occur at low frequencies below the measured range and are estimated from the simulated model curve. The estimated relaxation time for the ionic layer ($t_2$) and bulk ($t_1$) was found to be lower than that for thin films reported previously (Table S4).[39] The values of the relaxation times for the different halide samples correlate with the respective ionic sizes.[41,42] Further, we have calculated ionic mobility ($\mu_i$) using the Nernst−Einstein relation ($\mu_i = qD_{ion}/kT$), the ionic density $N_{ion}$ obtained from $\sigma_i$ (Table S2) and $\mu_i$ using the $N_{ion} = \sigma_i/q\mu_i$ relation. The halide ion density $N_{ion}$ shows a wide variation of $10^{16}$–$10^{22}$ cm$^{-3}$, the highest ion concentration for Cl (up to ~$10^{22}$ cm$^{-3}$), and the lowest for I (up to ~$10^{17}$ cm$^{-3}$).

**4.7. Dark I−V Study Using Space Charge Limited Current (SCLC).** The dark SCLC I−V measurement is a steady-state technique used to probe the effect of ionic defect states, interfacial defects, and grain boundaries on charge carrier transport in semiconductors by injecting charge carriers without the influence of photogenerated charge carriers.[34,43] Dark I−V characteristics can be subdivided into three regions: (1) at low bias, on carrier injection, the curve follows Ohm's law ($I \sim V^n$, n

= 1) with the measured conductivity essentially due to the background charge carriers; (2) at intermediate bias, the injected (free) charge carrier concentration becomes greater than the background charge carrier concentration. As the injected carriers fill up the trap states (defects), the current increases sharply with a non-Ohmic slope, $I \sim V^{n>2}$. The slope of the I–V curve is affected by the type and extent of defects in the material. This is called the trap state or trap-filled region; (3) at high bias, after all trap states are occupied, the space charge buildup leads to saturation in the current response, with $I \sim V^{n=2}$. This is called Child's region.

The size of crystals in methylammonium lead halide ($MAPbX_3$) perovskite materials can have a significant impact on the SCLC I–V response. In general, larger and well-formed crystals tend to have lower defect densities and fewer grain boundaries that act as carrier trapping centers. Consequently, the charge carriers (electrons and holes) have longer mean free paths and longer diffusion lengths, resulting in an enhanced carrier mobility. We compare below the morphologies and measurement geometries of our samples (polycrystalline film, polycrystalline pellet, and single crystal) that could affect the SCLC results.

The grain (crystallite) sizes of the 1 mm thick polycrystalline pellets were about 1–10 $\mu$m as measured by SEM and comparable to those in the 5 $\mu$m thick polycrystalline films. The corresponding grain/crystallite sizes of the 1 mm thick single crystals were not measured but are expected to be much larger. The SCLC measurements were performed across the thickness of pellets/crystals and across a 5 × 5 $mm^2$ area in the plane of the film. The larger grain/crystallite size in the I and Cl single-crystal devices was typically associated with higher values of saturation current as compared to the polycrystalline samples. This would be expected due to the lower defects/grain boundaries with correspondingly large mean free paths and greater mobility of charge carriers. The Br single-crystal samples had suffered some degradation due to exposure before dielectric and SCLC electrical characterization could be performed, resulting in increased granularity (confirmed by SEM) and hence showed a similar response as the corresponding polycrystalline samples.

Figure 13 shows the room-temperature measurements of dark I–V SCLC on $MAPbI_3$ single-carrier (electron/hole)-type devices in the forward bias mode. The polycrystalline thick film (~ 5 $\mu$m) single-carrier devices used a Ag anode and FTO cathode. Between the FTO cathode and perovskite absorber, the electron-assisted device (Figure 13a) used Al:ZnO as the electron transporting layer, while the hole-assisted device (Figure 13b) used PEDOT–PSS as the hole transporting layer. The single-crystal and polycrystalline pellet (~1 mm thick) devices used Ag as both anode and cathode (Figure 13c,d).

The contact between the Ag metal and $MAPbX_3$ (X = I, Br, Cl) semiconductor has an injection barrier of about 0.4, 0, and 1.3 eV for I, Br, and Cl, respectively (Figure S13). It, however, exhibits an Ohmic response at low biasing due to the mobile halide ions that act as internal (self) dopants due to quasi-Fermi-level alignment at the metal–semiconductor interface, allowing easy injection and extraction of carriers. Consequently, the SCLC I–V response is linear at low biasing and at high biasing (Child's region) becomes quadratic due to current saturation.

The devices exhibit a wide variation in their SCLC I–V response in the trap state region. Some curves exhibit a distinct trap state region at intermediate bias (Figures 13, S9, and S10).

The Cl thick films (Figure S9d) and polycrystalline pellets (Figure S10c) show only a small signature of a trap state region, while in the Br single crystal (Figure S10b), no trap states can be observed ($N_{ion} = N_{trap}$).[44] The I and Cl single-crystal devices (Figures 13d and S10d) show signatures of a distribution of trap states with different energy levels that are filled at different voltages, as suggested by recent simulation reports.[45]

The dark SCLC I–V curves were used to calculate the conductivity ($\sigma$), trap states ($N_{trap}$), carrier mobility ($\mu$), carrier concentration ($N_c$), and carrier diffusion length ($L_D$). The results are listed in Tables S5 and S6.

The conductivity was calculated from the slope of the I–V curve in the Ohmic region using eq 10

$$\sigma = \frac{I}{VL} \quad (10)$$

where I is the current, V is the applied voltage, and L is the thickness of the device (~ 5 $\mu$m for thick film, 1 mm for polycrystalline and single crystal). The estimated conductivity for the devices ($1–10^{-5}$ S/m) had a very large spread but was much higher than that reported previously ($10^{-7}$ S/m) for thin film and single-crystal devices.[34,46,47]

The density of trap states ($N_{trap}$) was calculated from the slope (>2) of the I–V curve in the intermediate bias region, corresponding to the onset of filling of the trap states by the injected carriers (eq 11)

$$N_{trap} = \frac{2\varepsilon_o \varepsilon V_{TFL}}{qL^2} \quad (11)$$

where $V_{TFL}$ is the trap-filled voltage, L is the thickness of the device, $\varepsilon_o$ is the permittivity of free space, $\varepsilon$ is the dielectric constant, and q is the elementary (electron/hole) charge. The thick film devices had an estimated trap state density of ~$10^{12}$ to $10^{13}$ $cm^{-3}$, which is lower than the $10^{14}$ to $10^{17}$ $cm^{-3}$ reported previously for thin films. The polycrystalline and single-crystal devices had an estimated trap state density of ~$10^9$ to $10^{10}$ $cm^{-3}$. The trap densities calculated for our samples are lower than other PV absorber materials such as polycrystalline-Si ($10^{13}–10^{14}$ $cm^{-3}$),[47,48] CdTe ($10^{11}–10^{13}$ $cm^{-3}$),[49] CIGS ($10^{13}$ $cm^{-3}$), rubrene ($10^{16}$ $cm^{-3}$),[50] and pentacene ($10^{14}$ to $10^{15}$ $cm^{-3}$).[51]

The carrier mobility ($\mu_n$) of the devices was determined from the values of current density $J_D$ and applied voltage V at the start of the high bias (Child's) region, shown in Figures 13, S9, and S10, where the approximately quadratic (n = 2) dependence obeys the Mott–Gurney (M-G) law. Equation 12 was used for the polycrystalline and single-crystal devices. Equation 13, using a dimensionality parameter $\xi_2$ modified for quasi 2-dimensional films, was used for the thick film devices[43,52,53]

$$\mu = \xi_1 \frac{J_D L^3}{\varepsilon \varepsilon_o V^2} \text{(bulk)} \quad (12)$$

$$\mu = \xi_2 \frac{J_D L^2}{\varepsilon \varepsilon_o V^2} \text{(film)} \quad (13)$$

where $\xi_1 \left(= \frac{8}{9}\right)$ and $\xi_2 \left(= \frac{7}{5}\right)$ are the dimensionality constants, $J_D$ is the current density, V is the applied voltage, $\varepsilon_o$ is the permittivity of free space, $\varepsilon$ is the dielectric constant, and L is the thickness of the device.[43,53] The results are listed in Tables S5 and S6.

The charge carrier mobility values in these materials are known to exhibit wide variation, ranging from 1 to $10^3$ $cm^2$/V–s,

depending on the device morphology (thin films, polycrystalline, single crystal), material synthesis conditions, and measurement techniques used (Hall, SCLC, PL quenching, electrical conductivity by THz optical pumping, transient microwave conductivity, etc.).[16] The mobility for the Cl and Br thick film devices was in the range of $10^2-10^3$ cm$^2$/V−s, while that for I thick films was much lower (<5 cm$^2$/V−s). This could be due to a number of factors discussed as follows: (i) the larger I$^-$ ion possesses a much greater polarizability, which produces a large internal field to counter the external applied bias, and leads to lower charge carrier mobility.[21] (ii) The use of an inert (FTO) anode along with Ag cathode leads to the formation of a redox shuttle of the halide ion across the absorber (Figure 14a), which leads to charge carrier (electron/hole) capture and reduced mobility.[54] This effect would be much more pronounced for the larger I ions moving through 5 $\mu$m thick films. The thick film single-carrier devices exhibit intrinsic semiconductor behavior, with the hole-assisted devices having mobility values smaller than those of electron-assisted ones. This is typical of semiconductors, in which the degeneracy at the top of the valence band leads to interband scattering and reduces the mobility.[55]

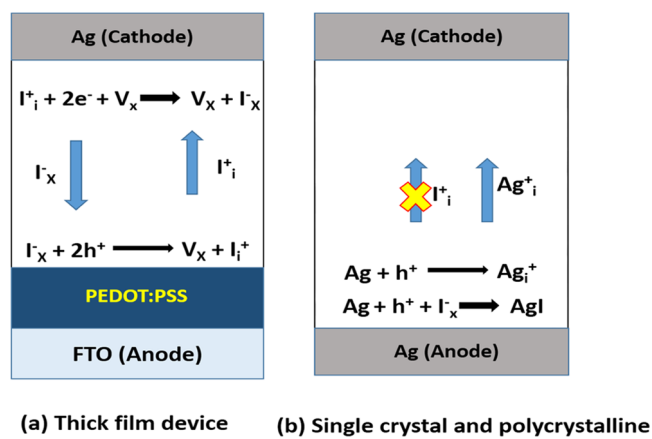

**Figure 14.** (a) Voltage-induced halide redox shuttle in thick film device with an inert FTO anode and an active Ag cathode; (b) suppression of halide redox shuttle by active Ag anode and cathode in polycrystalline and single-crystal devices.

In contrast to the thick film devices, for polycrystalline and single crystalline single-carrier devices, the mobility of the I samples was much higher at ∼$10^2$ cm$^2$/V−s, while that of the Cl and Br samples decreased sharply to ∼1 cm$^2$/V−s. This could be due to a number of factors as stated below: (i) the use of Ag for both cathode and anode, instead of the FTO cathode with a charge transporting layer, forms an electrode−absorber interface with a different injection barrier as compared to the transporting layer (Figure S13). (ii) The Ag anode acts to suppress halide oxidation and therefore the mobile halide ion redox shuttle observed in the thick film devices (Figure 14b).[54] This reduces the level of charge carrier capture by mobile halide ions and enhances their mobility. The redox shuttle suppression could explain the increase in the mobility of polycrystalline and single crystalline I devices as compared to thick films (Tables S5 and S6). The corresponding Cl and Br device mobilities, however, show a sharp decrease compared to that of the thick film devices. A possible explanation is that the large device thickness allows a much greater overall charge carrier capture due to the Coulomb attraction by the much more electronegative Cl and Br ions as compared with I ions. This would result in a sharp fall in the carrier lifetimes and mobilities for the former.[44]

The above transport behavior is also supported by the I−V hysteresis curves (Figures S11 and S12). The ∼5 $\mu$m thick film single-carrier device of MAPbI$_3$ exhibits a significant prebias effect (current rectification) and high hysteresis due to the larger size and higher polarizability of the I$^-$ ion, which is not observed in the corresponding Br and Cl devices. In contrast, the ∼1 mm thick polycrystalline and single-crystal devices exhibit a significant prebias effect of ∼10 V for Cl, ∼ 2 V for Br, and ∼1 V for I, which can be attributed to the greater Coulomb charge capture mentioned above.

The carrier concentration $N_c$ was calculated from the conductivity (eq 10) and mobility (eqs 12 and 13)

$$N_c = \frac{\sigma}{q\mu} \quad (14)$$

The estimated values of $N_c$ ($10^8-10^{15}$ cm$^{-3}$) are comparable with the previously reported studies of SCLC and Hall measurements on MAPbX$_3$ (X= I, Br, Cl).[15,56,57]

The diffusion length ($L_D$) was calculated using the recombination lifetime ($\tau$) obtained from TCSPC measurements (Figure 7) and the estimated mobility extracted from SCLC measurements (eqs 12 and 13)[58]

$$L_D = \sqrt{D\tau} \quad (15)$$

where $D\left(=\frac{k_B T}{q}\mu\right)$ is the diffusion coefficient. The calculated diffusion lengths $L_D$ were in the range of ∼0.1 to 6 $\mu$m. The wide variation observed in diffusion lengths in bulk and films could be due to grain boundary effects, contact resistance, and variation in carrier concentration as mentioned earlier. However, the values are comparable to other reports on monovalent cation-based hybrid perovskites.[34,35]

## 5. CONCLUSIONS

We have demonstrated a facile low-temperature growth of MAPbX$_3$ (X = I, Br, Cl) single crystals. The XRD and SEM/HR-TEM characterizations show that the crystals are of high quality and single phase in composition. We have prepared devices of different sample morphologies, spin-coated thick films, polycrystalline, and single crystalline, and used structural, optical, dielectric, and SCLC measurements to study the mixed nature of ionic−electronic charge transport in these materials.

The samples of all morphologies had low defect concentrations, as evidenced by the small absorptive energy loss in optical measurements. The mixed ionic−electronic behavior of the MAPbX$_3$ samples was studied experimentally and by simulations using an ionic−electronic model. The dielectric measurements showed a clear signature of the trapping effects of halide ion migration (X$^-$) and grain boundaries on the electronic charge carrier transport. The single-crystal samples showed a lower defect capture cross section ($\sigma_-$) compared to polycrystalline samples. The $\sigma_-$ was found to be proportional to the simulated halide ion density $N_{ion}$, which varied in the range of $10^{16}-10^{22}$ cm$^{-3}$ depending on the halide ion. The trap density ($N_{trap}$) obtained from SCLC measurements was found to be similar for a given morphology for different halide samples, and the single/polycrystalline samples had $N_{trap}$ ∼ $10^9-10^{10}$ cm$^{-3}$, which is much lower than previously reported. However, the simulated ionic density $N_{ion}$ varied greatly with the halide ion, leading to the variation of the effective density of the mobile halide ions ($N_{eff} = N_{ion} - N_{trap}$) over a large range ($10^2–10^{12}$ cm$^{-3}$). The single-crystal I samples showed a low capture cross section ($\sigma_- \sim 10^{-16}$ cm$^{-2}$), high mobility ($\mu \sim 196$ cm$^2$/V-s), and large diffusion length ($L_D \sim 6$ $\mu$m).

The results of our study on MAPbX$_3$ absorbers show that despite the presence of high ionic density, the nonradiative energy loss and carrier trapping can be suppressed and transport properties enhanced by (i) using single crystals to reduce grain boundary effects, in combination with (ii) interface engineering in the form of suitable electrodes to neutralize the mobile ions.

## ■ ASSOCIATED CONTENT

### ⓈSupporting Information

The Supporting Information is available free of charge at https://pubs.acs.org/doi/10.1021/acsaelm.3c00513.

Synthesis of MAPbX$_3$ (X = Br, Cl) polycrystalline powder and single crystals; powder XRD spectra for polycrystalline samples; XRD Rietveld refinement; SEM micrographs; HR-TEM images and histogram of nanoparticle size distribution; extracted UV−vis, PL, and TRPL data; dielectric response of polycrystalline and single crystalline MAPbX$_3$ (X = Br, Cl); dielectric parameters; defect capture cross section; simulated impedance curve; simulated relaxation time and $R_iC_i$ circuit parameters; I−V curves for thick film, polycrystalline, and single-crystal devices; transport parameters for thick film, polycrystalline, and single-crystal single-carrier devices; I−V hysteresis curves for thick film; polycrystalline and single-crystal devices; and HUMO−LUMO energy level diagram (PDF)

## ■ AUTHOR INFORMATION

**Corresponding Author**

    **Asad Niazi** − *Department of Physics, Jamia Millia Islamia, New Delhi 110025, India;* ⓘ orcid.org/0000-0003-2421-8357; Email: aniazi@jmi.ac.in

**Authors**

    **Mohd Warish** − *Department of Physics, Jamia Millia Islamia, New Delhi 110025, India*

    **Gaurav Jamwal** − *Department of Physics, Jamia Millia Islamia, New Delhi 110025, India*

    **Zara Aftab** − *Department of Physics, Jamia Millia Islamia, New Delhi 110025, India*

    **Nidhi Bhatt** − *Department of Physics, Jamia Millia Islamia, New Delhi 110025, India*

Complete contact information is available at:
https://pubs.acs.org/10.1021/acsaelm.3c00513

**Notes**

The authors declare no competing financial interest.

## ■ ACKNOWLEDGMENTS

Mohd Warish acknowledges the DST-Inspire Fellowship funded by the Department of Science and Technology, New Delhi. The authors acknowledge the Central Instrumentation Facility, JMI, and Inter University Accelerator Center (IUAC), New Delhi, for providing the characterization facilities for the study. The authors thank Dr. Pumlianmunga, Department of Physics, JMI, New Delhi, for providing the thermal vapor deposition and I−V measurement facility. The authors thank Dr. Motiur Rahman Khan, Dept of Physics, JMI, for his valuable suggestions regarding analysis of the experimental results.